\documentclass[graybox]{svmult}

\usepackage{type1cm}        
\usepackage{makeidx}         
\usepackage{graphicx}        
\usepackage{multicol}        
\usepackage[bottom]{footmisc}

\usepackage[numbers]{natbib}
\usepackage{newtxtext}       %
\usepackage[varvw]{newtxmath}  

\newcommand{\ZT}[1]{\textquotedblleft#1\textquotedblright}%
\usepackage{amsmath}

\makeindex             

\begin{document}

\title*{Neural networks consisting of DNA}
\author{Michael te Vrugt}
\institute{Michael te Vrugt \at Institut f\"ur Physik, Johannes Gutenberg-Universit\"at Mainz, 55128 Mainz, Germany, \email{tevrugtm@uni-mainz.de}}
%
%
\maketitle

\textit{This book chapter will appear in: M. te Vrugt (Ed.), Artificial Intelligence and Intelligent Matter, Springer, Cham (2025)}

\hspace*{1cm}

\abstract{Neural networks based on soft and biological matter constitute an interesting potential alternative to traditional implementations based on electric circuits. DNA is a particularly promising system in this context due its natural ability to store information. In recent years, researchers have started to construct neural networks that are based on DNA. In this chapter, I provide a very basic introduction to the concept of DNA neural networks, aiming at an audience that is not familiar with biochemistry.}

\section{Introduction}

Recent years have seen an increasing amount of work (some of which is also covered in this book) on implementing machine learning methods in physical systems, and the concept of intelligent matter \cite{KasparRvdWWP2021} is closely related to this idea. While many approaches of this type employ electronic, magnetic, or photonic systems, it is in principle a relatively natural idea to use soft and biological matter as a basis for physical neural networks. After all, artificial neural networks are inspired by the brain, and the brain is a soft matter system. 

DNA, the carrier of genetic information, naturally suggests itself for such approaches. It is a soft matter system that has evolved specifically for the purpose\footnote{Provided one can speak about things like \ZT{purposes} of natural objects in a scientific context, see Ref.\ \cite{Hundertmark2024} for a discussion.} of storing and processing information. Moreover, there is an established tradition of using DNA for performing artificial computational tasks in the framework of \textit{DNA computing} \cite{Adleman1994}. Therefore, a number of authors \cite{KimHW2004,QianWB2011,XiongZZCXLWFP2022,GenotFR2013,EvansOWM2024} have explored the possibility of using DNA as a basis for artificial neural networks.

In this chapter, I will provide an introduction to the topic of DNA-based neural networks, in particular the approach presented in Ref.\ \cite{CherryQ2018}. The presentation will be on a relatively elementary level, without assuming prior knowledge in biochemistry (and thereby aiming at an audience coming from physics or computer science interested in such approaches). After a brief overview over the biology of DNA (Section \ref{bfb}) and DNA computing (Section \ref{dnac}), I will present the basic ingredients of a DNA neural network, namely winner-take-all networks (Section \ref{wta}) and DNA gates (Section \ref{lc}). Then, I will present two approaches to DNA-based artificial intelligence, namely winner-take-all networks operating with DNA (Section \ref{adnn}) and DNA reservoir computing (Section \ref{drc}). Finally, I discuss some advantages and disadvantages of this approach (Section \ref{da}). I conclude in Section \ref{s}. In my presentation, will follow Refs.\ \cite{BolsoverHSWW2004} (in Section \ref{bfb}), \cite{KumarN2023} (in Section \ref{dnac}), \cite{QianW2011b} (in Section \ref{lc}), \cite{CherryQ2018} (in Sections \ref{wta} and \ref{adnn}), and \cite{GoudarziLS2013} (in Section \ref{drc}).

\section{\label{bfb}Biochemistry for beginners}
I have promised that this chapter does not require prior exposure to biochemistry, and consequently I will start with a brief introduction to what DNA is.

DNA (short for \textit{deoxyribonucleic acid}) is a biological macromolecule that stores the genetic information. A macromolecule is (surprise!) a large molecule. Macromolecules often consist of many repeated subunits (\textit{monomers}), in this case they are referred to as \textit{polymers}. DNA is a polymer whose subunits are nucleotides. Each nucleotide consists of a base, deoxyribose (a sugar), and a phosphate group. The deoxyribose of one nucleotide binds to the phosphate group of the next one (this is referred to as \textit{phosphodiester bond}).

What matters most for our purposes is the base, of which there exist five types: adenine (A), guanine (G), cytosine (C), thymine (T), and uracil (U). In DNA, one only finds adenine, guanine, cytosine, and thymine. These can bind to each other via hydrogen bonds (\textit{base pairing}). More specifically, adenine always binds to thymine and cytosine always binds to guanine (the bonding between G and C is a bit stronger). DNA is therefore typically found in a helix structure consisting of two \textit{strands} (chains of nucleotides). The nucleotides within a strand are bound together by the phosphodiester links, nucleotides in different strands are held together via base pairing. Since each base only binds to one specific other base, knowing the composition of one strand allows to infer the composition of the other one (the strand are \textit{complementary}). For example, if the first strand is ACCCGAT, the second one has to be TGGGCTA.

In protein synthesis, DNA is converted into proteins (biological macromolecules that perform a variety of functions). This occurs in several steps. First (\textit{transcription}), DNA is copied to ribonucleid acid (RNA), specifically to so-called messenger RNA (mRNA). RNA is similar to DNA, with differences being that it is usually single-stranded, that it contains ribose rather than deoxyribose, and that it uses uracil rather than thymine as the complementary base to adenine. At the \textit{ribosome}, the mRNA is then translated to proteins (\textit{translation}) according to the \textit{genetic code}, where a sequence of three nucleotides corresponds to one amino acid. For example, if UUC corresponds to the amino acid phenylalanine and AGG to the amino acid arginine, then the sequence UUCAGG tells the ribosome to put phenylalanine and arginine together. (Usually these sequences are of course longer.) This flow of genetic information is summarized in the \textit{central dogma of molecular biology} \cite{Crick1958,Crick1970}.

\section{\label{dnac}DNA Computing}
The first DNA computer was realized in 1994 by \citet{Adleman1994}, who used it to solve the so-called \ZT{directed Hamiltonian path problem}. Here, the aim is to find, for directed graph and a given starting and end point, a path that visits each node exactly once. This is a classical problem in computer science, since it is easy to pose, but very difficult (\ZT{NP-hard}) to solve. Adleman's procedure allowed, exploiting the parallelism of DNA computing, to solve this problem with a procedure where the number of steps grew only linearly with the number of vertices. This is a notable efficiency since the number of potential solutions increases combinatorially with the number of vertices, as a consequence of which traditional computing approaches do not achieve such a scaling \cite{BaumgardnerEtAl2009}. A discussion of the many developments and applications of DNA computing that emerged after Adleman's work can be found in review articles \cite{KumarN2023,Ezziane2005,MaCZHT2021,XuT2007}. Of particular interest in the context of this book is the development of intelligent systems based on DNA (see Ref.\ \cite{Ezziane2005} for a review). 

Some major advantages of DNA in the context of computing are \cite{KumarN2023}
\begin{itemize}
    \item \textbf{Parallelism:} DNA computers can perform a large number of tasks in parallel, which can lead to extremely good performances compared to modern supercomputers. For instance, Adleman's DNA computer already had a performance of 100 Teraflops.
    \item \textbf{Storage:} DNA allows for extremely efficient data storage, requiring just 1 cubic nanometer for one bit of information. All the datat hat humanity has generated by 2025 could be stored in the size of a ping-pong ball, in a manner that makes the stored data easy to maintain and to copy \cite{IonkovS2021}.
    \item \textbf{Energy Efficiency:} DNA computers, being based on chemical reactions, do not require any electricity and are very energy efficient compared to traditional computers.
\end{itemize}
However, there are also significant disadvantages \cite{KumarN2023}:
\begin{itemize}
    \item \textbf{Accuracy:} The biochemical processes involved in DNA computers are prone to errors, with the error probability increasing exponentially with the number of operations. 
    \item \textbf{Resources:} DNA computers are fairly difficult to handle, requiring familiarity with molecular biology and biochemical experiments. Carrying out these experiments requires human interventions in most steps. 
\end{itemize}
(These statements about advantages and disadvantages refer to what the technology is in principle capable of, currently it is in its infancy and is not easily available to practitioners who desire, say, high parallelism, as alternative to electronic approaches.)


\section{\label{wta}The winner takes it all}
The DNA-based implementation of neural networks discussed in this chapter is based on so-called \textit{winner-take-all computation} \cite{Maass2000}. Here, the basic idea is that neurons compete with each other. They inhibit other neurons while activating themselves. Thereby, it can be ensured that only the neuron with the largest input stays active, while all the other ones become inactive. A CMOS implementation of a winner-take-all function was proposed in Ref.\ \cite{LazzaroRMM1988}. How exactly this architecture can be implemented in a DNA system, and how input signals can be encoded in DNA, will be discussed in Section \ref{adnn}.

In the context of neural networks, a winner-take-all computation can connect the penultimate and final layer of a neural network, which have the same number of neurons. Let us denote the $i$th neuron in the penultimate layer by $s_i$ and the $i$th neuron in the final layer by $y_i$. The connection is set up in such a way that $y_i$ is one if $s_i$ is the largest value in the penultimate layer and zero otherwise.

How is this achieved? Thinking of one layer of a neural network as a vector $\vec{x}$ with components $x_i$, the next layer is constructed by multiplying the vector by a matrix and then applying a nonlinear function to each component of the resulting vector. Denoting the elements of the weight matrix by $w_{ij}$, the $j$th component of the vector that results from applying the weight matrix to the penultimate layer is 
\begin{equation}
s_i = \sum_{j}p_{ij}
\label{matrixmultiplication}
\end{equation}
with the products
\begin{equation}
p_{ij} = w_{ij} x_j.
\end{equation}
The nonlinear function is then one that selects the largest component of the vector $\vec{s}$:
\begin{equation}
y_i = 
\begin{cases}
1 \text{ if }s_i > s_j \forall j \neq i,\\
0 \text{ otherwise.}
\end{cases}
\label{nonlinear}
\end{equation}

Such a setup allows the network to have memory, which is encoded in the weights, and to solve pattern recognition tasks. Suppose the network has been trained to remember two patterns corresponding to the vectors $\vec{w}_1$ and $\vec{w}_2$ (for example, two different letters) with components $w_{1i}$ and $w_{2i}$. Then, it receives as an input a vector $\vec{x}$ corresponding to some pattern, and it is supposed to figure out whether it is more similar to $\vec{w}_1$ or to $\vec{w}_2$. Then, one has to simply multiply the vector $\vec{x}$ by the matrix $w_{ij}$ to get the vector $\vec{s}$ with components $s_i$ given by Eq.\ \eqref{matrixmultiplication}. Whether or not the pattern is more similar to $\vec{w}_1$ or $\vec{w}_2$ can then be figured out by checking whether $s_1 = \vec{w}_1\cdot\vec{x}$ or $s_2 = \vec{w}_2 \cdot\vec{x}$ is larger. Consequently, a network of this form requires $nm$ weights to remember $m$ $n$-bit patterns.

\section{\label{lc}DNA gates}

A key technique in this context is \textit{toehold-mediated strand displacement} \cite{YurkeTMSN2000,ZhangW2009} (see Ref.\ \cite{SimmelYS2019} for a review). This process involves a single-stranded DNA (the \textit{input}) and a double-stranded DNA, whose strands are referred to as \textit{gate} and \textit{output}. One strand of the double-stranded DNA (the gate) has an overhanging piece (the \textit{toehold}) that the input can bind to. Via branch migration (a process by which one DNA strand is exchanged for another, see Ref. \cite{HsiehP1995} for an introduction), the input strand then gradually starts binding to the gate strand and thereby replaces the output strand, which is thereby released. This process is more likely to start if the toehold is longer, and consequently, the toehold length can be used to control the reaction rate \cite{QianW2011}. Specifically, the reaction rate increases exponentially with the toehold length \cite{ZhangW2009}. (In practice stochastic fluctuations are of course likely to be relevant here, what exactly their influence is has not been systematically investigated.)

\begin{figure}
\includegraphics[scale=0.4]{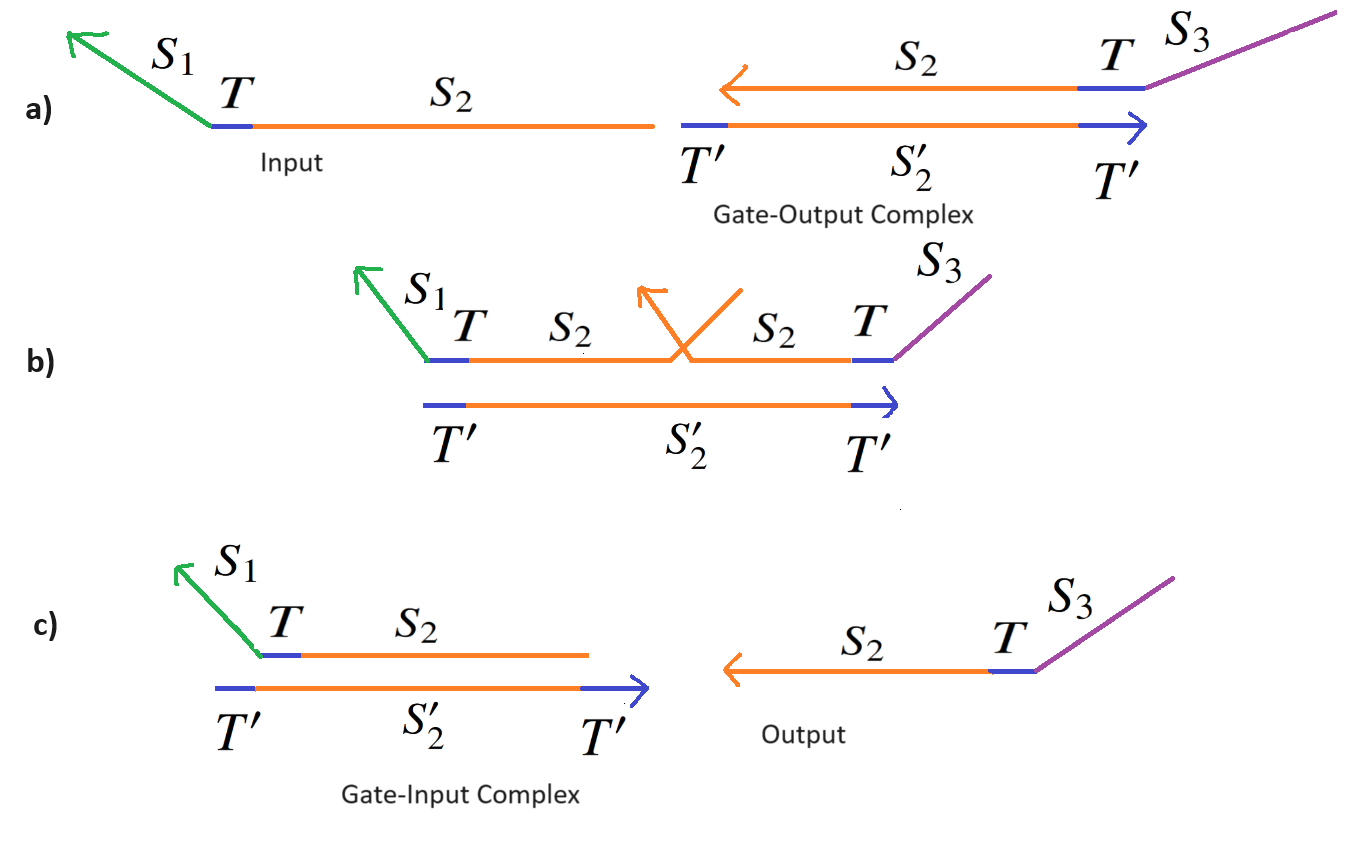}
\caption{Visualization of toehold exchange. (A similar illustration can be found in Ref.\ \cite{QianW2011b}.)}
\label{toeholdexchange}
\end{figure}

This process is visualized in Fig.\ \ref{toeholdexchange}. (A more complex illustration can be found in Ref.\ \cite{QianW2011b}.) Fig.\ \ref{toeholdexchange}(a) shows the initial configuration. The input strand consists of three segments, labelled $S_1$, $T$, and $S_2$. Here, $S_1$ and $S_2$ are \textit{recognition domains}, which are relatively long (15 nucleotides), and $T$ is the \textit{toehold domain}, which is shorter (5 nucleotides). The other necessary ingredient is the \textit{gate-output-complex}, which consists of two strands (gate and output) that are bound to each other. The gate has a recognition domain $S_2'$ with a toehold domain $T'$ before and after it, while the output has two recognition domains $S_2$ and $S_3$ separated by a toehold domain $T$. Here, a prime indicates a complementary base sequence (e.g., if $T=$AGGAT, then $T'$=TCCTA). Due to the DNA base-pairing rules (see Section \ref{bfb}), the segments $S_2'$ and $T'$ bind to the segments $S_2$ and $T$ of the output, respectively.

Since the gate has two toeholds and the output has only one, the gate-output complex has a free toehold $T'$. This toehold is complementary to the free toehold $T$ of the input. As a consequence, the input binds to the gate-output complex, leading to a complex consisting of three DNA strands. Now, although the recognition domain $S_2'$ of the gate is currently bound to the output, it could by the DNA base-pairing rules 
equally well bind to the recognition domain $S_2$ of the input. Gradually, via branch migration, the recognition domain of the output ceases to bind to the $S_2$ recognition domain of the output and instead binds to the corresponding domain of the input (as shown in Fig.\ \ref{toeholdexchange}(b)). In the final state, shown in Fig.\ \ref{toeholdexchange}(c), the gate is now bound to the input, forming a \textit{gate-input complex}, while the output has been released. 

To summarize, what can be achieved via this process is to convert an input signal (which is a DNA strand) into a desired output signal (which is also a DNA strand) with a certain rate. In particular, this method allows to realize so-called \textit{seesaw gates} \cite{QianW2011,QianW2011b}. In addition to input and output, also \textit{fuel} (a third type of strand) is present. In the simple setup presented above, input strands release the output strands from the gates and are then no longer available for further reactions. If fuel is present, however, the fuel strand can (via toehold exchange) free the input strand and thereby make it available for further reactions. Thereby, a small amount of input can, provided enough fuel is available, in principle trigger the release of an arbitrary amount of output. The fuel therefore acts as a catalyst \cite{QianW2011b}.

\section{\label{adnn}A DNA neural network}
We have now assembled all ingredients that are necessary to understand a simple winner-take-all neural network based on DNA, namely the one realized by \citet{CherryQ2018}. What it is supposed to do is to recognize handwritten numbers based on the general approach discussed in Section \ref{wta} (and it turns out that the DNA neural network presented in Ref.\ \cite{CherryQ2018} is in fact able to recognize handwritten numbers -- which is a standard benchmark task for artificial neural networks). The input vector (reprenting the image) is multiplied by a weight matrix (matrix multiplication consists of multiplication of the elements of the vector with scalar weights and summation over the results), afterwards the entries of the resulting vector are compared, the largest component is increased and the other ones are eliminated to generate a definite output.

\begin{figure}
    \centering
    \includegraphics[scale=0.3]{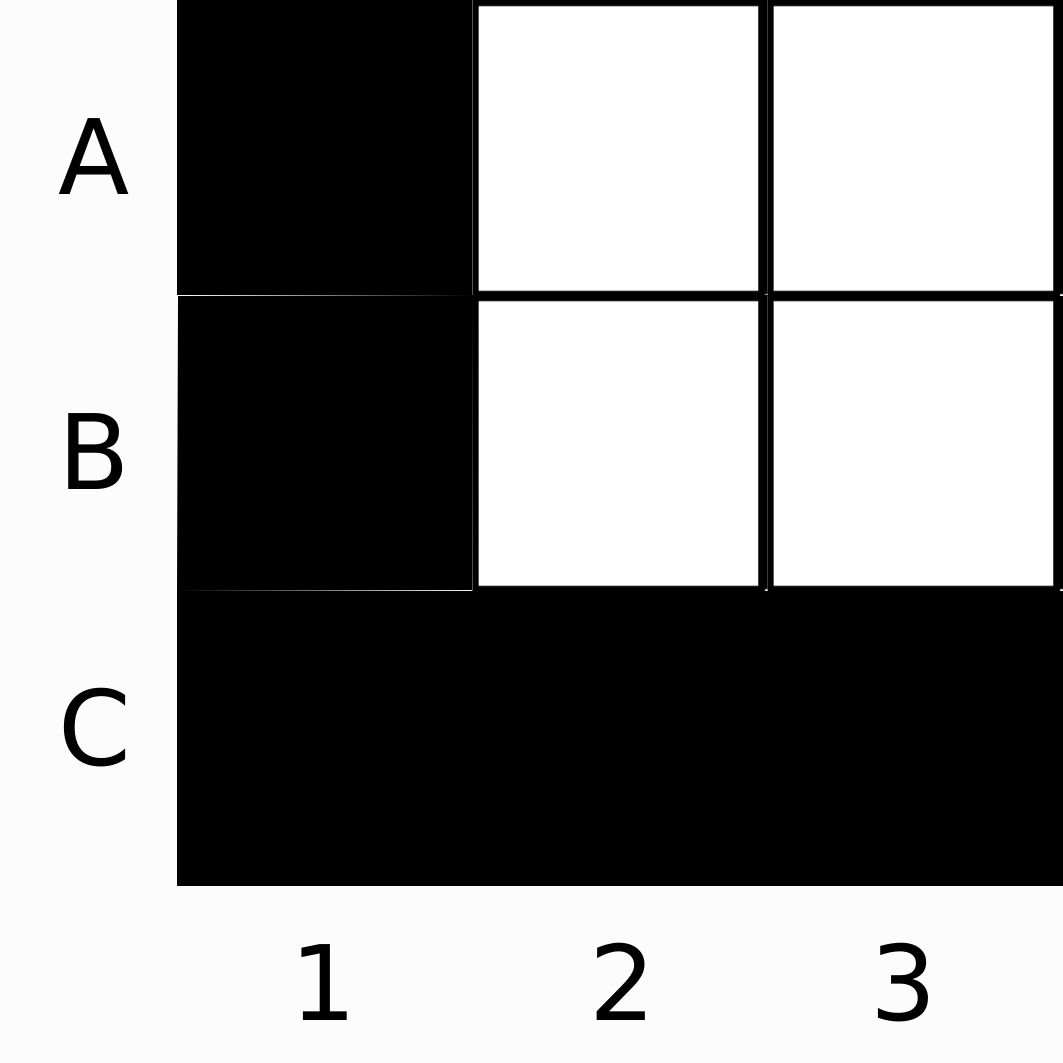}
    \caption{Feeding a hand-written letter (L) as an input signal in the form of DNA into the system. The nine pixels are represented by nine distinct DNA molecules. An L is encoded by  the presence of A1, B1, C1, C2, and C3. (A similar illustration can be found in Fig. 1b of Ref.\ \cite{CherryQ2018} and in Fig. 2b of Ref.\ \cite{teVrugt2024}.)}
    \label{fig:digit}
\end{figure}

First, we need to find a way of encoding, for example, a hand-written digit in the form of DNA strands, to allow for processing via chemical reactions. Consider as an example a nine-pixel image of the letter L, as shown in Fig. \ref{fig:digit}. We choose a set of nine distinct DNA molecules (A1, A2, A3, B1, B2, B3, C1, C2, C3) that represent the nine pixels of the image. In the case of the letter L, five of the nine pixels (A1, B1, C1, C2, C3) are black and the others white. Consequently, we can encode the letter L as an input by putting the DNA molecules A1, B1, C1, C2, and C3 into the DNA computer. (In practice one would instead use 100 pixels represented by 100 distinct molecules \cite{CherryQ2018}, but that does not change the general idea.)

Second, we have to find a way of implementing the matrix multiplication \eqref{matrixmultiplication} and the nonlinear function \eqref{nonlinear}  via DNA. Let us start with the matrix multiplication. Importantly, one is concerned here with a binary input, i.e., all $x_i$ are either zero or one). In the context of number recognition, where the $x_i$ correspond to pixels, this might for example indicate whether the pixel represented by $x_i$ is black or white. The entries of the weight matrix $w_{ij}$, on the other hand, are analogue numbers. What one requires here is thus a reaction that converts $w_{ij}$ to $p_{ij}$ if and only if $x_i$ is present, regardless of the exact concentration of $x_i$. This is achieved by a catalytic cycle involving $x_i$, $w_{ij}$ and fuel. The species $x_i$ binds to a gate, corresponding to $w_{ij}$, that then releases a product $p_{ij}$. Afterwards, the fuel releases the $x_i$ strand from the gate, such that it can trigger releases of $p_{ij}$ elsewhere. Provided that a sufficient amount of fuel is present, this process produces a concentration of $p_{ij}$ corresponding to the concentration of $w_{ij}$, but only if $x_i$ is present.

Summation is relatively straightforward: Recall that seesaw gates allow to convert a certain input signal (a chemical substance, more specifically a DNA strand) into an output signal (another chemical substance, more specifically another DNA strand). For summation, one therefore simply requires gates that convert all species $p_{ij}$ to the same species $s_j$. The concentration of $s_j$ is then the sum of the concentrations of the $p_{ij}$.

The next step is to determine which of the $s_i$ is the largest one. Given that the $s_i$ correspond to the concentrations of chemical species, this is achieved by pairwise annihilation reactions, implemented by annihilator molecules. If two strands $s_i$ and $s_j$ bind to an annihilator, it splits into two waste molecules that are not able to participate in further reactions. After a while, only the species that had the largest concentration will remain. Then, a signal-restoration reaction that converts $s_i$ to the output $y_i$ takes place. This is again implemented catalytically via seesawing reactions -- if $s_i$ is present, in any concentration, then the fuel ensures that a sufficient amount of $y_i$ is produced, but no production of $y_i$ takes place if no $s_i$ is present. Thereby, one has found a biochemical implementation of the nonlinear function \ref{nonlinear}. Finally, the output is converted to a fluorescent signal.

Given that the neural network consists of chemicals floating around in a test tube, it is not immediately clear why these calculations should take place in the desired order (as they do in a traditional neural network implemented in a computer). For most steps, this is ensured by the fact that a reaction can only take place if the reactants are present. The products of one step (one reaction) are the reactants for the next one. The exception are the annihilation and restoration reactions, where the reactants (the $s_i$) are the same. Here, the desired order is ensured by different reaction rates: The annihilation reaction is very fast compared to the restoration reaction. Thereby, the $s_i$ quickly eliminate each other until only one is left, the survivor's concentration then gets very slowly increased again. Specifically, the toeholds of the annihilator have two extra nucleotides, which due to the exponential scaling of the reaction rate with the toehold length lead to a reaction rate increased by a factor of about 100.

\section{\label{drc}DNA reservoir computing}
Finally, I will discuss a different approach to DNA-based machine learning. This approach, proposed in a theoretical study by \citet{GoudarziLS2013}, uses \textit{reservoir computing} \cite{Jaeger2001,MaassNM2002}. See the introductory chapter on this topic in this volume for an explanation of how reservoir computing works and the chapters by Kathy L\"udge/Lina Jaurigue, Atreya Majumdar/Karin Everschor-Sitte, and Julian Jeggle/Raphael Wittkowski in this volume for other implementations. 

In a nutshell, reservoir computing employs a dynamical system (the reservoir) that is driven by an input signal. The response of the system then serves as the input for a neural network with a single layer (the \textit{readout layer}) that converts this response into the output layer.  This readout layer is the only part of the system that is changed during the training process. Since the reservoir does not have to be changed, the reservoir can also be a physical system (for example one consisting of DNA). It is widely assumed\footnote{It is not really clear at present to what extend this is actually the case, see the chapter on reservoir computing in this book.} that it is helpful if the reservoir operates close to criticality, such that the dynamics is rich enough to allow for interesting things to be read out from it (as opposed to, say, a system where all trajectories approach a certain stationary state irrespective of the input).

\citet{GoudarziLS2013} suggest that such a rich transient dynamics can be found in a network consisting of coupled chemical oscillators, realized by a microfluidic chamber containing different interacting DNA species. As a starting point, they use a network proposed by \citet{FarfelS2006}. Substrate molecules enter a reaction chamber and bind to gate molecules, by which they are then converted to product molecules. The presence of the product molecules inhibits the reaction of substrates and gates. Moreover, the products flow out of the chamber with a certain rate. This gives rise to a reaction-inhibition cycle leading to sustained oscillations of the product concentrations. Denoting these by $P_i$, and moreover the substrate concentrations by $S_i$, the substrate influx rates per volume by $S_{i,\mathrm{in}}$, the gate concentrations by $G_i$, the efflux rate by $e$, the volume of the reactor by $V$, the well-mixed fraction of the reactor by $h$, and the reaction rate by $\beta$, this behavior is described by the dynamical system \cite{GoudarziLS2013}
\begin{align}
\dot{P}_1&=h\beta S_1(G_2 - P_3)-\frac{e}{V}P_1,\\
\dot{P}_2&=h\beta S_2(G_2-P_1)-\frac{e}{V}P_2,\\
\dot{P}_3&=h\beta S_3(G_3-P_2)-\frac{e}{V}P_3,\\
\dot{S}_1&=S_{1,\mathrm{in}}-h\beta S_1(G_1-P_3) - \frac{e}{V}S_1,\\
\dot{S}_2&=S_{2,\mathrm{in}}-h\beta S_2(G_2 - P_1) - \frac{e}{V}S_2,\\
\dot{S}_3&=S_{3,\mathrm{in}}-h\beta S_3(G_3 - P_2) - \frac{e}{V}S_3.
\end{align}
As detailed in Ref.\ \cite{GoudarziLS2013}, a linear stability analysis of this dynamical system allows to find the parameter region in which it exhibits oscillatory behavior. A sustained oscillation exists for $h\beta (S_1S_2S_3)^\frac{1}{3}/2 - e/V = 0$.

The influx rate $S_{1,\mathrm{in}}$ is used to drive the system, this is the input layer. The reservoir state is specified by the concentrations of the various substances. Of particular importance here are the product concentrations $P_1$, $P_2$, and $P_3$. It is assumed that one can read out the reservoir state using a fluorescent probe. The known values of $P_1$, $P_2$, and $P_3$ are then fed into the readout layer (which is trained via linear regression.). The readout layer can be implemented on a computer, or as a part of the physical system. In total, one thus has a neural network consisting of a microfluidic DNA chamber and a single-layered neural network (for example on a computer) that receives an input signal in the form of an influx rate and returns an output signal that is obtained by feeding the measured product concentrations in the DNA chamber into the readout layer.

\section{\label{da}Advantages and disadvantages of this approach}
While it is certainly impressive that test tubes filled with some chemicals can be used for handwritten number recognition, it should of course be noted that this is certainly not the most efficient approach if the recognition of handwritten numbers is our primary goal. If the numbers are primarily a proof of principle, what then can these methods be useful for?

DNA neural networks require the input signal to have the form of a DNA strand. In general, this is a disadvantage since converting general input signals to DNA is quite an effort. This aspect can, however, turn into an advantage in contexts where the input signal takes the form of DNA strands (or at least that of biomolecules) anyway. This will primarily be the case in biomedical applications of neural networks. Suppose, for example, that a neural network has been trained to recognize genetic dispositions for a certain disease. If this network is implemented in DNA form, then one could just take a DNA sample from the patient, put it into a test tube, and then see a glowing test tube indicating that the gene one looks for is (or is not) present.

A further aspect to note here are the time scales involved here. Photonic neural networks (see the chapters by Kathy L\"udge/Lina Jaurigue and by Lennart Meyer/Rongyang Xu/Wolfram Pernice) have the attractive feature that they operate with the speed of light, i.e., they are extremely fast. This can certainly not be said about DNA computers. Their computational speed depends on how fast the chemical reactions take place, which can be of the order of many minutes. An advantage, in contrast, is that DNA-based approaches are well suited for parallelization. These (dis)advantages are familiar from DNA computing in general.

Moreover, there can be contexts where having a neural network that operates slower can actually be an advantage.\footnote{Thanks to Raphael Wittkowski for bringing this to my attention.} A good example would be a network that processes temporal input signals, as is required, e.g., in speech recognition. In this case, it is advantageous if the system's dynamics takes place on roughly the same time scales as the input signal rather than being significantly faster. Consider reservoir computing, where the employed physical system possesses a fading memory, as an example -- if, after the second word of a sentence to be processed, all memory of the first word has already vanished, the system cannot process the sentence as a whole. A particular advantage of DNA neural networks in this context is that the speed at which the reactions take place (and thereby the speed at which the network operates) can be tuned by the experimenter, namely by changing the lengths of the toeholds (see Section \ref{lc}).

Finally, from a conceptual or pedagogical point of view, it should be noted that (if we ignore the details of the biochemical processes involved here) the neural network presented in Section \ref{adnn} has an extremely simple architecture. It is therefore possible to explicitly model and understand how each entry of an intermediate state vector and each weight contributes to the final output of the network. Therefore, this approach helps to teach basic concepts of artificial intelligence to high school and undergraduate university students, a teaching concept based on this idea is outlined in Ref.\ \cite{teVrugt2024}. Moreover, studying networks of this type allows to get a better intuition for how neural networks generally come to the conclusions they come to, a goal that, in the framework of \ZT{explainable artificial intelligence} \cite{GunningSCMSY2019} has motivated several works studying simple network architectures in other contexts (including biomolecular ones \cite{BraghettoOB2023}). For the same reason, the template of DNA neural networks can provide inspiration for other attempts to implement artificial intelligence in soft matter systems (see, for example, the chapters by Hartmut L\"owen/Benno Liebchen, Julian Jeggle/Raphael Wittkowki, Giovanni Volpe, and Jannes Freiberg/Roshani Madurawala in this volume), and it is a helpful case study for philosophical discussions of physical computing (see the chapter by Luis Lopez in this volume). 
\section{\label{s}Summary}
In this chapter, I have provided a brief introduction to neural networks consisting of DNA, using as an example the winner-take-all network proposed in Ref.\ \cite{CherryQ2018}. The input data is provided as a DNA strand and is processed via biochemical reactions. On this basis, it is possible to recognize handwritten digits using DNA. Moreover, I have briefly discussed a proposal for DNA-based reservoir computing \cite{GoudarziLS2013}. Such approaches constitute a promising starting point for the development of intelligent matter based on biological materials, and might also find applications in, for instance, medical contexts where input data is already present in a biochemical form. 

\section{Acknowledgements}
This work was funded by the Deutsche Forschungsgemeinschaft (DFG, German Research Foundation) in the framework of SFB 1551; Project No. 464588647 and SFB 1552; Project No. 465145163.

\end{document}